\documentstyle[12pt,aasms4]{article}

\lefthead{Sil'chenko}
\righthead{NGC 7331}

\begin{document}

\title{NGC 7331: the galaxy with the multicomponent central region}
\slugcomment{Partly based on observations collected with the 6m
telescope at the Special Astrophysical Observatory (SAO) of the
Russian Academy of Sciences (RAS).}

\author{O. K. Sil'chenko\altaffilmark{1}}
\affil{Sternberg Astronomical Institute, Moscow, 119899 Russia\\
       Isaac Newton Institute, Chile, Moscow Branch\\
     Electronic mail: olga@sai.msu.su}

\altaffiltext{1}{Guest Investigator of the RGO Astronomy Data Centre}

\begin{abstract}

We present the results of the spectral investigation of the regular
Sb galaxy NGC~7331 with the Multi-Pupil Field Spectrograph of the
6m telescope. The absorption-line indices H$\beta$, Mgb, and
$<\mbox{Fe}>$ are mapped to analyse the properties of the
stellar populations in the circumnuclear region of the galaxy.
The central part of the disk inside $\sim 3\arcsec$ (200 pc) -- or
a separate circumnuclear stellar-gaseous disk as it is distinguished
by decoupled fast rotation of the ionized gas -- is very metal-rich,
rather young, $\sim 2$ billion years old, and its solar
magnesium-to-iron ratio evidences for a very long duration of the
last episode of star formation there. However the gas excitation
mechanism now in this disk is shock-like. The star-like nucleus
had probably experienced a secondary star formation burst too: its
age is 5 billion years, much younger than the age of the circumnuclear
bulge. But [Mg/Fe]=+0.3 and only solar global metallicity imply that
the nuclear star formation burst has been much shorter than that
in the circumnuclear disk. The surrounding bulge is rather old,
9--14 billion years old, and moderately metal-poor. The rotation of
the stars and gas within the circumnuclear disk is axisymmetric
though its rotation plane may be slightly inclined to the global
plane of the galaxy. Outside the circumnuclear disk the gas may
experience non-circular motions, and we argue that the low-contrast
extended bulge of NGC~7331 may be triaxial.

\end{abstract}


\section{Introduction}

The nearby spiral galaxy NGC~7331 the main parameters of which
are given in Table 1 represents a challenge for the
observers interested in structure and dynamics of disk galaxies.
It has been observed many times, both photometrically and
spectroscopically; but results of each new approach contradicted
often to results of previous ones. Many years ago Bosma (\cite{bosma})
trying to give a general description of NGC~7331 noted that
according to Sandage (\cite{ha}) who saw spiral arms as close to
the center as at $r\approx 6\arcsec$ NGC~7331 is a disk-dominated
galaxy; however the deep photograph reported by Arp and Kormendy
(\cite{ak72}) revealed a presence of prominent extended bulge.
The problem of the bulge role in NGC~7331 is not solved yet despite
numerous photometric studies. Boroson (\cite{bor}) estimated a
bulge-to-disk ratio as 1.10 by analysing a major-axis surface
brightness profile. Kent (\cite{kent}) analysed two-dimensional
CCD images and proposed a method of bulge-disk decomposition
based on different ellipticities of the bulge and disk isophotes;
but he noted that this method is inapplicable to NGC~7331 because
in this galaxy the isophotes of the bulge and of the disk
demonstrate the same (!) ellipticity. As a result, he decomposed
only the major-axis profile and derived $B/D=0.66$ and the disk
with a central hole. Among recent studies, von Linden et al.
(\cite{germ}) have published a surface brightness profile in the
$I$--band extended up to $R\approx 100\arcsec$ and have concluded
that a de Vaucouleurs' bulge dominates over the whole radius range
under consideration, whereas Prada et al. (\cite{espls}) analysing
together $I$-- and $K$--images have reached the best fit with the
compact bulge, having effective radius of $\sim 10\arcsec$, and
two exponential disk components with different characteristic
scales which meet at $R\approx 100\arcsec$. So the situation with
the morphological characteristics of the bulge in NGC~7331
remains to be uncertain. A similar uncertainty exists relating
to a dynamical status of the bulge in this galaxy: Prada et al.
(\cite{espls}) have found the bulge to counter-rotate with
respect to the stellar disk basing on the long-slit observations
in the \ion{Ca}{2}\,IR triplet spectral range, but Mediavilla et
al. (\cite{esptig}) who observed the central part of NGC~7331
with a panoramic fiber spectrograph in two spectral ranges, near
Mgb and near \ion{Ca}{2}\,IR, do not agree stating that the
bulge in NGC~7331 corotates the stellar and gaseous disks.

\begin{table}
\caption[ ] {Global parameters of NGC~7331}
\begin{flushleft}
\begin{tabular}{lc}
\hline\noalign{\smallskip}
Type (NED) & SA(s)b \\
$R_{25}$, kpc (LEDA) & 22.5 \\
$B_T^0$ (LEDA) & 9.27  \\
$M_B$ (LEDA) & --21.41 \\
$V_r(radio) $ (LEDA) & 821 $km\cdot s^{-1}$   \\
Distance, Mpc (Hughes et al. 1998) & 15.1  \\
Inclination (LEDA) &  $70^o$  \\
{\it PA}$_{phot}$ (LEDA) &  $171^\circ$  \\
\hline
\end{tabular}
\end{flushleft}
\end{table}

Another question which is an object of discussion for several
years: is there a "dead quasar", or supermassive black hole,
in the center of NGC~7331? We have begun this discussion ten
years ago. Afanasiev et al. (\cite{asz89}) examining the
major-axis profile of line-of-sight velocities of the ionized gas
have found that central $\pm 2\arcsec$ (the scale which is
comparable to our spatial resolution) are kinematically
decoupled by fast solid-body rotation. We have checked the
axisymmetric character of the gas rotation and have concluded
that this fast rotation is caused by a compact mass concentration of
order of $5 \cdot 10^8 \, M_\odot$. To see if a "dead quasar" may be
in the center of NGC~7331, Bower et al. (\cite{betal}) have
studied major-axis and minor-axis profiles of stellar velocities.
Though the angular rotation velocity of stars in the center of
NGC~7331 has been found to be as high as that of the ionized gas,
namely, about of 500 km/s/kpc, the very central part of the
stellar rotation curve has not appeared to be so distinct.
Moreover, the stellar velocity dispersion has not a sharp
maximum in the center, as it is the case, for example, in M~31
where the presence of the supermassive black hole in the
center is proved (\cite{dr88,k88}). On the contrary, in NGC~7331
$\sigma (r)$ along the minor axis seems to demonstrate a local
minimum at $r\approx 0\arcsec$. So Bower et al. (\cite{betal})
have concluded that the mass of the central black hole, if
exists, must be less than $5 \cdot 10^8 \, M_\odot$. However,
the problem of the black hole presence in the nucleus of NGC~7331
has not been closed: Cowan et al. (\cite{crb}) have reported
a detection of unresolved nuclear radio source in this galaxy
which has appeared to be more luminous by a factor of 3--4
than the famous Sgr~A in the center of our Galaxy, and recently
Stockdale et al. (\cite{src}) have claimed an existence of the
nuclear X-ray source in NGC~7331 -- they treat it as a massive
black hole. If it really exists, its dynamical implications must be
re-looked for more carefully.

When Afanasiev et al. (\cite{asz89}) have detected a compact mass
concentration in the center of NGC~7331, we have not made a definitive
choice between the supermassive black hole and a compact dense
stellar subsystem, such as a central star supercluster or
circumnuclear stellar disk. Moreover, later we have undertaken a
two-dimensional spectrophotometry of the central region of NGC~7331
and have found that its unresolved nucleus is chemically distinct:
its magnesium index is much higher than that of the nearest bulge
(\cite{sil92}). As the kinematically decoupled nucleus in NGC~7331
has been found to be also chemically distinct, we would like to think
it to be a separate stellar subsystem. So a more careful investigation
of stellar population properties in the nucleus and in the circumnuclear
region of NGC~7331 must help to clarify the structure and evolution
of its center and of the entire galaxy. We report our observations and
other data which we use in Section~2. The brief description
of the ionized-gas morphology in the center of NGC~7331 is given
in Section~3. Radial variations of the stellar population age
and metal abundances are analysed in Section~4, and the
kinematics of ionized gas and stars in the region under
consideration is discussed in Section~5. Section~6 presents
our conclusions and a brief discussion of our results.

\section{Observations}

The observations of NGC~7331 which results are presented in this
paper have been undertaken with the Multi-Pupil Field Spectrograph
(MPFS, \cite{afetal90}) of the 6m telescope of the Special Astrophysical
Observatory of the Russian Academy of Sciences in 1996.

The red spectral range, 6300--6900~\AA, has been exposed on August 19,
1996, during 55 minutes, under the seeing $FWHM \approx 2\farcs 4$.
The MPFS was equipped
with CCD $520 \times 580$; we registered simultaneously 95 spectra
from an area of $19\arcsec \times 12\farcs 6$, each spectrum
corresponding to a spatial element of $1\farcs 58 \times 1\farcs 58$.
The strongest emission line in the circumnuclear region,
[NII]$\lambda$6583, has been measured to construct two-dimensional
velocity field of the ionized gas; we fitted its profile by a
single Gaussian.

The blue-green spectral range, 4100--5600~\AA, has been exposed on
October 9, 1996, during 40 minutes, under the seeing $FWHM=1\farcs 6$.
The sky has been exposed separately during 20 minutes, properly
normalized and subtracted from the galaxy frame. The MPFS
was equipped with CCD $1040 \times 1160$; we registered
simultaneously 128 spectra from an area of
$11\arcsec \times 21\arcsec$, each spectrum corresponding to a
spatial element of $1\farcs 33 \times 1\farcs 33$. This spectral
range includes a lot of strong absorption lines, so we have used
this exposure to calculate absorption-line indices H$\beta$, Mgb,
Fe5270, and Fe5335 in the well-known Lick system (\cite{woretal}).
We have checked our consistency with the Lick
measurements by observing stars from their list (\cite{woretal})
and by calculating the absorption-line indices for the stars in the same
manner as for the galaxies. The indices calculated for 9 stars are
coincident with the data tabulated in (\cite{woretal}) in average within
0.05~\AA. The exposure for the galaxy has been taken long enough
to provide a signal-to-noise ratios of $\sim$ 60 in the nucleus
and $\sim$ 10--15 near the edge of the frame; the corresponding random
error estimations made in the manner of Cardiel et al. (\cite{cgcg})
range from 0.2~\AA\ in the center to 0.6--0.7~\AA\ for the individual
spatial elements at the outermost points. To keep a constant level of
accuracy along the radius, we summed the spectra for the galaxies in
concentric rings centered onto the nuclei and studied the radial
dependencies of the luminosity-weighted properties of stars by
comparing the observational index values to those from the
synthetic models of old stellar populations of Worthey (\cite{worth94})
and Tantalo et al. (\cite{tantalo}). We estimate the mean accuracy of
our azimuthally-averaged indices as 0.1--0.2~\AA. Besides, we
have cross-correlated each elementary spectrum with the spectrum
of the late-type star ADS~15470 (the brighter component) and
have obtained two-dimensional velocity field for the stellar
component of the central region of NGC~7331.

The reciprocal dispersion during these observations was 1.6~\AA\ per
pix, and the spectral resolution varied slightly over the field of
view from 3.5~\AA\ to 5~\AA. As a result, we estimate our accuracy
of the elementary line-of-sight velocities as 20--25 km/s. However,
our spectral resolution allowed to obtain only luminosity-weighted
mean line-of-sight velocitites; we were not able to separate
kinematical components similar to the
counterrotating bulge reported by Prada et al. (\cite{espls}).

To refine kinematical analysis, we have involved high-resolution
long-slit data from the La Palma Archive. The galaxy has been observed
on July 19, 1996, with the ISIS (red arm) equipped with CCD TEK
$1024 \times 1024$ at the William Herschel Telescope. The near-infrared
spectral range, 8360--8750~\AA, containing the strong CaII
absorption-line triplet, has been exposed: 60 min in the
$P.A.=172\arcdeg$ (major axis) and 30 min in the $P.A.=262\arcdeg$
(minor axis). The slit width was $0\farcs95$; the reciprocal dispersion
of 0.39~\AA\ per pix provided spectral resolution of 1.0~\AA. A star
HR~8656 (K0III) was observed during the same night; we have
cross-correlated the galaxy spectra row-by-row with the spectrum of
this star after binning by three pixels (with the final spatial
step of $1\arcsec$), sky subtracting and transforming into velocity
scale. The subsequent Gauss analysis of obviously multi-component
LOSVDs (line-of-sight velocity distributions) has allowed to extract
several kinematical components along the slit aligned with the major
axis of the galaxy.

The basic reduction steps -- bias subtraction, flatfielding,
cosmic ray hit removing, extraction of one-dimensional spectra,
wavelength calibration, construction of surface brightness maps --
have been fulfiled by using the software developed in the Special
Astrophysical Observatory (\cite{vlas}). To calculate the
absorption-line indices and to map them we have used our own
programs as well as the FORTRAN program of Dr.~Vazdekis.

\section{Morphology of the Ionized Gas Distribution in the Center
of NGC~7331}

Fig.~1 shows raw observational data -- so called "data cube" -- in
the red spectral range. One can see immediately that a characteristic
ratio of emission lines, $H\alpha$-to-[NII]$\lambda$6583, indicates
a LINER-like excitation inside the central region, roughly
$16\arcsec \times 10\arcsec$, but a rather strong present star
formation already in $11\arcsec$ to the east (along the minor axis)
from the nucleus. This fact contradicts to the claims of Keel
(\cite{keel}) who observed the spectra in $10\arcsec$ to the north,
south, and east (!) from the nucleus and everywhere had found the
dominance of [NII]$\lambda$6583 over H$\alpha$. But his aperture,
$8\farcs 1$, was perhaps too large to distinguish between the central
shock-excited gaseous disk and nearby star-forming sites which are
concentrated in a ring with a deprojected radius of $45\arcsec$
(or $15\arcsec$ on the sky plane along the minor axis). This ring
is presented best of all in the recent paper of Smith (\cite{smith}).
It looks quite identical at different wavelengths: on the
H$\alpha$+[NII] map from Pogge (\cite{pogge}), on the 15$\mu$m map
from Smith (\cite{smith}), on the Nobeyama CO (1-0) map from Tosaki
and Shioya (\cite{jap}), and on the 20\,cm radio
continuum map from Cowan et al. (\cite{crb}). Obviously, this multiple
coincidence together with H$\alpha$/[NII]$\lambda 6583 > 2$ in our
Fig.~1 proves that this ring contains sites of intense present star
formation, including formation of massive stars. If we accept the
surface-brightness profile decomposition from Boroson (\cite{bor})
or from Baggett et al. (\cite{bag}), this star-forming ring is
located well inside the bulge-dominated area.

The nature of the very central region is not so obvious. Fig.~2
presents isophotes of the [NII] emission-line surface brightness
distribution obtained with the MPFS. They are elongated and repeat
approximately the shape of continuum isophotes. The consistent
results are obtained by Mediavilla et al. (\cite{esptig}) (see their
Fig.~9); even an asymmetry "north-south" is the same. Keel (\cite{keel})
also mentioned a gaseous disk with the radius of $2\arcsec$ having the
diffuse extension up to $10\arcsec$ aligned in the $P.A.=165\arcdeg$,
close to the line-of-nodes orientation. Interestingly, the similar
elongated structure, with the major-axis diameter of
$7\arcsec - 9\arcsec$, is present on the 15$\mu$m map of Smith
(\cite{smith}), but the sense of the north-south asymmetry is opposite
with respect to the [NII] map. Probably, the explanation of this
asymmetry by the dust concentration to the south from the nucleus
given by Mediavilla et al. (\cite{esptig}) is correct. There are no
either HI (\cite{begeman}) nor CO (Tosaki \& Shioya \cite{jap})
inside this circumnuclear structure; also it is undetected in radio
(Cowan et al. \cite{crb}). Therefore, there is no detectable present
star formation there; but the gas is probably shock-excited, and
the dust is warm. As the structure is aligned with the line of nodes,
we would like to treat it preliminarily as a circumnuclear gaseous disk.

\section{Stellar Population Properties in the Circumnuclear Region
of NGC~7331}

Figure~3 presents isolines of the surface distributions in the
central $9\arcsec \times 10\arcsec$ area for the green (5100~\AA)
continuum and three absorption-line indices, Mgb,
$<\mbox{Fe}>\equiv$ (Fe5270+Fe5335)/2, and H$\beta$.
The continuum isophotes show a strange asymmetry: within $3\arcsec$
from the center eastern halves of them seem to be more tightly
packed than their western halves. It looks like a some extracomponent
of light to the west from the nucleus. Interestingly, on larger
scales the sense of isophote crowding is opposite due to global
dust concentration on the western side of the galaxy
(Boroson \cite{bor}, Smith \cite{smith}). The magnesium-index
isolines are rather roundish; the distribution is strongly peaked
on the nucleus of the galaxy (NGC~7331 has a chemically distinct
nucleus, as we noted earlier, \cite{sil92}). But the surface
distribution of another metal-line index, $<\mbox{Fe}>$, is quite
different from that of Mgb and can partly explain an asymmetry
of the continuum isophotes. In Fig.~3c we clearly see a compact
Fe-rich disk shifted to the west from the nucleus (or do we see
only its western half?). A strong north-south asymmetry is present
too. If we assume that we see only south-western quarter of the
disk, its radius may be as large as $3\farcs5$. This size agrees
with the size of the central mid-infrared structure
(Smith \cite{smith}). But what is the most interesting thing, it is
the similar surface distribution of the hydrogen-line index
H$\beta$ (Fig.~3d): the circumnuclear Fe-rich disk is also
distinguished by the very prominent Balmer absorption line. Since iron
and hydrogen absorption lines in integrated spectra are unmatched
being produced by different groups of stars, the similarity
of the Fig.~3c and Fig.~3d may signify that the circumnuclear
Fe-rich stellar disk is rather young. Let us look at what state-of-art
diagnostics of the stellar population properties can tell us.

To compare our measurements to the stellar population models based on
summation (with some weights) of spectra of stars, we must made
corrections for the stellar velocity dispersion in galaxies which
broadens absorption lines and "degrades" a spectral resolution in
such way. We have calculated the correction by smoothing the spectra
of K0-K3 III giants from the list of (\cite{woretal}) which we have
observed and by measuring the absorption-line indices of the smoothed
spectra. We have found that the index H$\beta$ is quite insensitive
to the velocity dispersion when $\sigma_v$ remains to be less than
230 km/s; as for the metal-line indices, we have found the
correction to be 0.1~\AA\ for $\sigma_v$=130 km/s which is typical
for the central part of NGC~7331 (Bower et al. \cite{betal}).
Figures~4 and 5 contains the corrected indices.

The most popular models of Worthey (\cite{worth94}) allow to disentangle
age and metallicity of old stellar populations by confronting
some metal-line indices (e. g. Mgb, $<\mbox{Fe}>$, or
[MgFe]$\equiv (\mbox{Mgb} \, <\mbox{Fe}>)^{1/2}$) with the
Balmer-line index H$\beta$; but these models have been calculated
for the solar magnesium-to-iron ratio. To use the Worthey's
(\cite{worth94}) models, we must be sure that the stellar population
has solar magnesium-to-iron ratio. Figure~4 presents the diagram
(Fe5270, Mgb) where we compare azimuthally-averaged (in circular
rings centered onto the nucleus) index measurements in NGC~7331
with the models of Worthey (\cite{worth94}) for the solar
magnesium-to-iron ratio. Since the circumnuclear Fe-rich disk
seen in Figs.~3c and 3d is located asymmetrically with respect to
the nucleus and contributes only partially to the ring-integrated
estimates at $r=1\farcs 3$ and $r=2\farcs 6$, we have also
plotted individual-element measurements related to this disk
though they are less accurate than azimuthally-averaged ones.
One can see from Fig.~4 that the nucleus of NGC~7331 is surely
magnesium-overabundant. The Mg/Fe ratio for the azimuthally-averaged
measurements at $r=1\farcs 3$ and $r=2\farcs 6$ is obviously lower
than that for the nucleus, but is it solar or not, depends on the
stellar population age. Farther from the nucleus, at $r \geq 4$,
the moderate magnesium overabundance can also be seen. Interestingly,
the proper measurements of the circumnuclear Fe-rich disk lie
much higher that the azimuthally-averaged points: the disk has
the solar Mg-to-Fe ratio if it is as young as 2 billion years old
and is iron-overabundant if it is older.

Taking in mind all said above, let us try to determine mean ages
of stars in the nucleus and at different distances from the center.
Figures~5a, 5b, and 5c present various diagrams which may be useful
for this purpose. As the nucleus is magnesium-overabundant, the
models of Worthey (\cite{worth94}) are inapplicable for it; in Fig.~5a
we have plotted the calculations of Tantalo et al. (\cite{tantalo})
for [Mg/Fe]=+0.3. From comparison with these models on the diagram
(H$\beta$, $<\mbox{Fe}>$) one can see that the mean age of the
nuclear stellar population in NGC~7331 is 5 billion years, and
its global metal content, $Z$, is close to the solar value. Fig.~5b
shows also the models from Tantalo et al. (\cite{tantalo}), but
for the solar magnesium-to-iron ratio; the locations of the
azimuthally-averaged points at the $r=1\farcs 3$ and $r=2\farcs 6$
imply similarly a rather young age for the nearest neighborhood of
the nucleus. The elementary measurements related to the circumnuclear
Fe-rich disk, though well scattered, nevertheless all lie above
the model sequence with the age of 5 billion years. The next Fig.~5c
presenting the diagram (H$\beta$, [Mg\,Fe]) with the models of
Worthey (\cite{worth94}) calculated under [Mg/Fe]=0 confirms that
four individual points for the circumnuclear Fe-rich disk agree
with the age estimate of 2 billion years and the overall metallicity
at least twice the solar one. If we return now to Fig.~4, we should
conclude that the circumnuclear "Fe-rich" disk has indeed [Mg/Fe]=0
under the assumption of $T=2$ billion years.

The work of Tantalo et al. (\cite{tantalo}) proposes also a possibility
to quantify differences of stellar population properties basing on
the index differences. A set of three linear equations, connecting
$\Delta$[Mg/Fe], $\Delta \log Z$, and $\Delta \log T$ to the
$\Delta \mbox{Mg}_2$, $\Delta <\mbox{Fe}>$, and $\Delta \mbox{H}\beta$,
is written. We apply these equations to the differences between the
nucleus and the bulge; the bulge is safely taken at the following values
of radius, at $r=4\arcsec$, $5\farcs 3$, $6\farcs 7$, and $8\arcsec$,
namely, outside the circumnuclear "Fe-rich" disk but well inside
the star-forming ring. Solely, one must take in mind that the index
measurements at $r > 6\arcsec$ are twice less precise than the more
inner ones so they can be used mostly as a check.
Having performed the set of calculations,
we have obtained the parameter differences listed in Table~2.
They mean that the bulge is twice older and more metal-poor
by a factor of 2.5--4 than the nucleus. Surprisingly, the Mg/Fe
ratios are almost equal in the nucleus and in the bulge. When we
compare the absolute values of nuclear indices,
$\mbox{Mg}_2=0.229 \pm 0.005$, $<\mbox{Fe}>=2.66 \pm 0.22$,
and H$\beta = 1.83 \pm 0.19$, to the direct model
calculations of Tantalo et al. (\cite{tantalo}), we see that the
model with $Z=0.02$ (solar value), [Mg/Fe]=+0.3, and $T=5$ billion
years has consistent index values. Then the bulge stellar population
parameters are $Z=0.005-0.008$, [Mg/Fe]=+0.2, and $T=9-12$ billion
years. The latter age estimate agrees also with the positions
of bulge points in Fig.~5a.

\begin{table}
\caption[ ] {Stellar population parameter differences "bulge-nucleus"}
\begin{flushleft}
\begin{tabular}{lccc}
\tableline
Radius, \arcsec & $\Delta$[Mg/Fe] & $\Delta \log Z$ & $\Delta \log T$ \\
\tableline
4 &  -0.09  & -0.39 & +0.26\\
5.3 & -0.17 & -0.41 & +0.29 \\
6.7  & -0.16 & -0.62 & +0.46 \\
8    &  -0.07 & -0.58 & +0.27 \\
\tableline
\end{tabular}
\end{flushleft}
\end{table}

The analysis undertaken in this Section has allowed to identify three
quite different stellar structures within $8\arcsec$ ($\sim$ 600 pc)
from the center. The unresolved star-like nucleus is rather young,
$5 \cdot 10^9$ years old, strongly magnesium-overabundant and has
solar global metallicity. Farther out, the circumnuclear disk with
a radius of $\sim 3\arcsec$ is even younger, $\sim 2$ billion years
old, has the solar Mg-to-Fe ratio, and the global metallicity higher
than the solar one. The surrounding bulge is older than the nucleus
and the circumnuclear disk, namely, is 9--12 billion years old,
magnesium-overabundant and moderately metal-poor.

\section{Kinematics of Gas and Stars}

Two-dimensional line-of-sight velocity fields for stars and ionized
gas obtained with the MPFS are presented in Fig.~6. Both fields
look rather regular and show clear signs of rotation. However there
are some differences between velocity fields of stars and gas. As
we noted earlier (Afanasiev et al. \cite{asz89}), the gas rotation
curve has a sharp local maximum at $R\approx 2\arcsec$, and we see
a consequent feature to the north-west from the center in Fig.~6b
(it is somewhat farther than it must be due to a seeing quality
worse with respect to that of our long-slit observations).
Meantime the rotation velocity of stars rises continuously
with radius -- rapidly up to $R\approx 5\arcsec$ and more slowly
farther from the center (Bower et al. \cite{betal}, Prada et al.
\cite{espls}), and consistently, Fig.~6a has no closed isovelocities.

Two-dimensional velocity fields can allow to check an axisymmetric
character of rotation. If we have an axisymmetric mass distribution
and rotation on circular orbits, the direction of maximum central
line-of-sight velocity gradient (we shall call it "dynamical major
axis") should coincide with the line of nodes as well as the
photometric major axis; whereas in a case of triaxial potential
the isovelocities align with the principal axis of the ellipsoid,
and generally the dynamical and photometrical major axes diverge
showing turns with respect to the line of nodes in opposite senses
(e. g. \cite{mbe92}). In a simple case of cylindric (disk-like)
rotation we have a convenient analytical expression for the
azimuthal dependence of central line-of-sight velocity
gradient within the area of solid-body rotation:\\

\noindent
$dv_r/dr = \omega$ sin $i$ cos $(P.A. - P.A._0)$, \\

\noindent
where $\omega$ is a deprojected central angular rotation velocity,
$i$ is an inclination of the rotation plane, and $P.A._0$ is an
orientation of the line of nodes.

We have fitted the data presented in Fig.~6a by this formula and
have obtained for the gradients taken within $r \leq 2\arcsec$:\\

\noindent
$dv_r/dr$ = [33 cos $(P.A. - 183^\circ) - 2.5$]
$\mbox{km} \ \mbox{s}^{-1} \ \mbox{arcsec}^{-1}$ . \\

\noindent
The amplitude of the cosine curve, $\omega \, \sin i = 33 \,
\mbox{km} \ \mbox{s}^{-1} \ \mbox{arcsec}^{-1}$, agrees rather well
with the central slopes of the major axis long-slit cross-sections
obtained for the stellar component of NGC~7331 by Bower et al.
(\cite{betal}) and by Prada et al. (\cite{espls}). Moreover, it agrees
rather well with the rotation velocity of ionized gas within the
kinematically decoupled region: in Afanasiev et al. (\cite{asz89})
we reported an azimuthal dependence of the central line-of-sight
velocity gradients based on the long-slit [NII]$\lambda$6583
emission line measurements in four different position angles,
the best-fitted cosine curve formula for which was:\\

\noindent
$dv_r/dr$ = [38.4 cos $(P.A. - 175^\circ) + 1.3$]
$\mbox{km} \ \mbox{s}^{-1} \ \mbox{arcsec}^{-1}$ . \\

\noindent
The phases of these two cosine curves are close too; but together
they evidence for a marginal turn of the dynamical major axis with
respect to the line of nodes which has $P.A.=166\arcdeg - 167\arcdeg$
(Prieto et al. \cite{pretal}, von Linden et al. \cite{germ}).
It would be a signature of a triaxial potential, if the photometric
major axis turns in opposite sense, to lesser $P.A.$. But indeed
it turns in the same sense! Numerous photometric studies detected
a turn of the photometric major axis in NGC~7331 to $P.A.\approx
175\arcdeg - 183\arcdeg$ inside $R=6\arcsec - 8\arcsec$ (somewhat
different in different works). It is also known that this turn
is stronger at shorter wavelengths. For example, Prieto et al.
(\cite{pretal}) have measured the following orientations of the
isophote major axis at $R\approx 4\arcsec$:
$P.A._0 \approx 182\arcdeg$ in the $B$-band,
$P.A._0 \approx 179\arcdeg$ in the $V$-band, and only
$P.A._0 \approx 176\arcdeg$ in the $I$-band; the asymptotic
$P.A._0$ value at larger radii given by them is $166\arcdeg$.
Up to now the common point of view is that it is a dust effect
which must be wavelength-dependent. But in the center of NGC~7331
the dust is still visible in the $I$-band, however, the measurements
of $P.A._0$ in the $I$- and in the $K$-bands agree well (Prada et al.
\cite{espls}). There may be another explanation: if there is a blue
(young) misaligned stellar disk inside the red (older) bulge, the
measurements of combined isophotes should show a stronger turn
through the bluer filter. The coincidence of the dynamical and
photometrical major axes at $P.A.\approx 180\arcdeg$ reveals a
presence of inclined circumnuclear stellar disk, rather rapidly
rotating.

The long-slit observations made under higher spectral resolution
may help to clarify a dynamical structure of the central region of
NGC~7331. The similar observational data taken along the major axis
of the galaxy in 1992 August have allowed Prada et al. (\cite{espls})
to claim a presence of counterrotating bulge in NGC~7331. Here we
present the later results for the major-axis and minor-axis
cross-sections. Fig.~7 shows a direct view of the LOSVD calculated along
the major axis, and Figs.~8a and 8b -- the results of multi-component
Gauss analysis of the LOSVDs.

One can immediately see a difference between our Fig.~7 and the
analogous Fig.~3 of Prada et al. (\cite{espls}). The description
of the latter included only two kinematical components: fast-rotating
disk and retrograde bulge rotating more slowly and seen only
in the radius range of $5\arcsec - 15\arcsec$. Prada et al.
(\cite{espls}) claimed an absence of prograde bulge. Meantime even
a single glance at our Fig.~7 reveals two quite noticeable
prograde structures: a fast-rotating disk and a slower rotating
prograde bulge seen up to $\sim 35\arcsec$ from the center. The
retrograde component is seen too, but it is not so prominent as
it seemed to be in Fig.~3 of Prada et al. (\cite{espls}). Since
our long-slit data were obtained with the same equipment -- the ISIS,
red arm, of the WHT, -- and the template star is of the same
spectral type as that of Prada et al. (\cite{espls}), we can only
refer to our better spectral resolution -- 35 km/s (their spectral
resolution was reported as 52 km/s).

Fig.~8a displays the results of Gauss analysis of the LOSVD along
the major axis $P.A.=172\arcdeg$ together with stellar velocity
profiles derived without component separation, namely, together
with the recent data of Heraudeau \& Simien (\cite{hr}) obtained under
seeing of $2\farcs 5$ and with the simulated one-dimensional profile
calculated from our MPFS two-dimensional velocity field shown in
Fig.~6a. By analysing the long-slit data, we have extracted three
kinematical components. The first, fast-rotating component dominates
over the full range of radii. Since according to Prada et al.
(\cite{espls}) at $R \geq 5\arcsec$ its stellar velocity dispersion
does not exceed 70 km/s, we would treat it as a disk. The second
component, corotating with the first one and seen at $R \geq 5\arcsec$,
rotates much slower than the first one, so we would conclude that
it is an ordinary bulge; but it dominates nowhere, and that is very
strange for the early-type spiral galaxy with extended photometric
bulge. Interestingly, its nearly solid-body slow rotation up to
$R \approx 45\arcsec$ matches perfectly the rotation of the ionized gas
at $R > 2\arcsec$ (see Fig.~1 in our work, Afanasiev et al.
\cite{asz89}, or the lower resolution H$\alpha$ data in \cite{fr})
while the rotation curve of the dominant stellar component, disk,
diverges strongly with the gas rotation curve. Finally, the third
component counter-rotates with respect to the main stellar and
gaseous rotation; it is just the same component which was reported
by Prada et al. (\cite{espls}) as a "retrograde bulge". But there
is some ambiguity with this diagnosis: in their Conclusions Prada
et al. (\cite{espls}) stated that "the inner parts of the galaxy
consists of a boxy component, dominating the inner $5\arcsec$. It
shows position angle twisting, rotates retrograde to the rest of
the galaxy, and is rounder"; meantime the retrograde component is
seen only in the narrow radius range $5\arcsec < R < 20\arcsec$
according both to the data of Prada et al. (\cite{espls}) and ours!
Besides, its velocity dispersion is as low as that of the disk
(Prada et al. \cite{espls}), and it rotates faster than the prograde
bulge though slower than the disk. So it does not resemble a bulge;
it must be a rather flat structure.

The central $5\arcsec$ contains only one inseparable component; but
at $R \approx 5\arcsec$ it meets perfectly the disk component. As
it looks like a straight solid-body rotation curve, we would identify
it with a central part of the disk rotation profile. An increased
velocity dispersion inside $R\approx 5\arcsec$ which has been reported
by Prada et al. (\cite{espls}) is probably a result of adding
slower rotating, weaker components. However, we can conclude
that from a dynamical point of view the disk dominates over
the whole central region of NGC~7331.

\section{Discussion}

We can summarize our conclusions as follows. In the Sb galaxy NGC~7331
the stellar disk is a kinematical component dominating over the full
radius range. The bulge is less prominent though it can be traced up
to $R\approx 45\arcsec$ (3.3 kpc); its slow solid-body rotation
is very similar to the main rotation of the ionized gas. We confirm
also an existence of the counter-rotating stellar component in
the radius range of $5\arcsec - 20\arcsec$ (400 -- 1500 pc). The
central part of the disk inside $\sim 3\arcsec$ (200 pc) -- or
a separate circumnuclear stellar-gaseous disk as it is distinguished
by decoupled fast rotation of the ionized gas -- is very metal-rich,
rather young, $\sim 2$ billion years old, and its solar
magnesium-to-iron ratio evidences for a very long duration of the
last episode of star formation there. However the gas excitation
mechanism in this disk now is shock-like. The star-like nucleus
had probably experienced a secondary star formation burst too: its
age is 5 billion years, much younger than the age of the circumnuclear
bulge. But [Mg/Fe]=+0.3 and only solar global metallicity imply that
the nuclear star formation burst has been much shorter than that
in the circumnuclear disk.

Up to now two-dimensional mapping of the absorption-line equivalent
widths (or indices) is rarely used to investigate stellar population
properties in galaxies. Practically unique examples of such approach
are a long-slit combined study of the center of M~31 by Davidge
(\cite{dav}) and a detailed TIGER investigation of NGC~4594 by
Emsellem et al. (\cite{ems96}). The latter case appears to be
something similar to the case of NGC~7331: the magnesium-index
map of the center of NGC~4594 demonstrates a point-like peak in
the nucleus of the galaxy, and the iron indices are roughly
constant along the major axis up to the border of the area
investigated, $R\approx 5\arcsec$ (Emsellem et al. \cite{ems96}).
So the isolines of the iron indices in NGC~4594 present something
like the Fe-rich circumnuclear stellar disk found by us in NGC~7331.
Though the measurements of Mgb index in NGC~4594 are complicated by
a presence of the rather strong emission line [NI]$\lambda$5199
(Emsellem et al. \cite{ems96}) and though two examples are not
a statistics yet, perhaps, there exists some evolutionary sense
in different morphologies of magnesium- and iron-index surface
distributions.

An interesting problem is an origin of the counter-rotating component.
Usually counter-rotating stellar substructures are considered
as a result of merger. Some merger event could also provoke a
secondary star formation burst in the center of the galaxy in this
way producing the chemically distinct nucleus and, if the merger
was dissipative, the inclined circumnuclear disk. However, the
circumnuclear Fe-rich disk in NGC~7331 rotates in the same sense
as the rest of the galaxy and so cannot be genetically related to
the counter-rotating component. Recently we have found a
counter-rotating stellar component in the nearby spiral galaxy
NGC~2841 (\cite{we99}). NGC~2841 and NGC~7331 look almost twins:
the same morphological type, Sb, the same size and luminosity, the
same inclination. They were the first galaxies where a global ring-like
distribution of CO has been detected (\cite{ys}). Analysing the
major-axis long-slit cross-section of NGC~2841, we have found two
kinematical components in the bulge: strong prograde one which
dominates up to $R\approx 25\arcsec$ and weak retrograde one
(\cite{we99}). A set of other phenomena allow us to suggest an
existence of extended triaxial bulge almost aligned with the line
of nodes of the global disk in this galaxy, and we have thought
the counter-rotating stellar component to be an intrinsic property
of a slightly tumbling triaxial potential. Perhaps the similar
situation takes place in NGC~7331; the only difference is that in
NGC~7331 the disk dominates over the full radius range whereas
in NGC~2841 the bulge is more prominent. A hypothesis of the
bar presence in the center of NGC~7331 has already been proposed
by von Linden et al. (\cite{germ}) to explain a central depletion
of molecular gas in this galaxy. Besides, the French team (\cite{fr})
have constructed a global two-dimensional velocity field of the
ionized gas in NGC~7331 from their observations of H$\alpha$ emission
line with a scanning Fabry-Perot interferometer and have found a
strong large-scale turn of the isovelocities which is usually treated
as a signature of bar presence. Now we can add another argument
in favor of a triaxial potential aligned with the line of nodes:
inside $R\approx 40\arcsec$ the gas in NGC~7331 rotates more slowly
than the stellar disk, and it is valid both for the ionized gas
(\cite{retal65}, Afanasiev et al. \cite{asz89}, \cite{fr}) and for
the molecular gas (von Linden et al. \cite{germ}). Since the emission
lines are narrow (below the spectral resolution) and the gas seems
to be well-settled to the global plane of the galaxy
(Bosma \cite{bosma}, von Linden et al. \cite{germ}),
the only explanation can be non-circular rotation of the gas caused
by the triaxial potential of the bulge. The stronger response of the
gaseous disk to the triaxial potential of the bulge when compared to
the response of the stellar disk may be explained by a viscous nature
of the gas and by a significant self-gravitation of the massive stellar
disk of NGC~7331 which is a dominant dynamical component in this galaxy.
Interestingly, the solid-body part of the gas rotation curve, or a
zone of non-circular motions as we think, ends at
$R=40\arcsec - 45\arcsec$  -- exactly at the ring of molecular gas,
warm dust, and intense star formation which we have discussed in
Section~3. The configuration looks like an HII ring around a bar
which is often observed in classic barred galaxies. Probably, the
low-contrast triaxial bulge extends up to $40\arcsec \div 50\arcsec$
from the center in NGC~7331.

The last interesting question concerns a possible presence of the black
hole in the center of NGC~7331. In Afanasiev et al. (\cite{asz89}) we
argued that the fast decoupled axisymmetric rotation of the ionized
gas inside $R=2\arcsec$ proved strong mass concentration in the nucleus;
this mass concentration might be a black hole. But Bower et al.
(\cite{betal}) have noticed no decoupled rotation by studying a stellar
component; and what is the most important, they have not found a peak
of stellar velocity dispersion in the nucleus. How can we agree gas and
star behaviours? In Afanasiev et al. (\cite{asz89}), Zasov and
Sil'chenko (\cite{zs96}) we noted that a misaligned minibar in the
center of a galaxy may mimic a presence of kinematically decoupled
nucleus by increasing a visible central slope of the major-axis
velocity profile due to non-circular motions. But in NGC~7331 the
situation is more complex: the rotation of the gas and stars in the
very center is circular, and the bar effect (non-circular gas motions)
is felt only outside the zone of decoupled rotation. However, the
result is the same: the nucleus looks kinematically decoupled though
no point-like central mass concentration is required for this. By
summarizing, since Cowan et al. (\cite{crb}) and Stockdale et al.
(\cite{src}) have found the unresolved radio- and X-ray source in the
center of NGC~7331, it may be a black hole, but it cannot be
a {\it supermassive} black hole, like those in NGC~4594, NGC~3115,
and M~87, because its dynamical influence is nowadays indeterminate.
As supermassive black holes are now detected in some three dozens
galaxies, a correlation is found between black hole mass and
galactic spheroid luminosity. By using this relation taken e. g.
from Cattaneo et al. (\cite{bhmass}) one can try to estimate
a possible black hole mass in the center of NGC~7331. Unfortunately,
as we have noted in the Introduction, the bulge-disk decomposition
in NGC~7331 is ambiguous: if the bulge is large as Boroson (\cite{bor})
or Baggett et al. (\cite{bag}) reported, the black hole mass
may be $10^9\,M_\odot$, if it is small like that reported by
Prada et al. (\cite{espls}), the mean relation $M_{BH}$ vs. $L_{B,bul}$
implies $M_{BH}=10^8\,M_\odot$ for NGC~7331. If we take also into
account the large scatter of this relation (Cattaneo et al.,
\cite{bhmass}, give $\sigma = 0.74$) and the impression that
the $M_{BH}$ estimates for spiral galaxies lie all below the mean
relation defined mostly by ellipticals, the mass of the black hole
in the center of NGC~7331 may be as small as $3 \cdot 10^7\,M_\odot$.
This value contributes only several percents into the total mass
of the circumnuclear disk ($5 \cdot 10^8\,M_\odot$, Afanasiev et al.,
\cite{asz89}), so it may be undetectable in kinematical studies
of moderate spatial resolution like ours.

\acknowledgements
I thank the astronomers of the Special Astrophysical Observatory
Drs. V. L. Afanasiev, S. N. Dodonov, V. V. Vlasyuk, and
Mr. S. V. Drabek for supporting the observations at the 6m telescope.
I am also grateful to the graduate student of the Moscow University
A. V. Moiseev for the help in preparing the manuscript.
The 6m telescope is operated under the financial support of
Science Department of Russia (registration number 01-43).
During the data analysis I have
used the Lyon-Meudon Extragalactic Database (LEDA) supplied by the
LEDA team at the CRAL-Observatoire de Lyon (France) and the NASA/IPAC
Extragalactic Database (NED) which is operated by the Jet Propulsion
Laboratory, California Institute of Technology, under contract with
the National Aeronautics and Space Administration.
This research has made use of the La Palma Archive. The telescope
WHT is operated on the island of La Palma by the Royal
Greenwich Observatory in the Spanish Observatorio del Roque de los
Muchachos of the Instituto de Astrofisica de Canarias. The work
was supported by the grant of the Russian Foundation for Basic
Researches 98-02-16196, by the grant of the President of Russian
Federation for young Russian doctors of sciences 98-15-96029
and by the Russian State Scientific-Technical
Program "Astronomy. Basic Space Researches" (the section "Astronomy").

\newpage

\figcaption{An example of raw data cube: small pieces of spectra
containing emission lines H$\alpha$ and [NII]$\lambda$6548,6583
are displayed on the gray-scaled continuum image. $P.A.(top)=97\arcdeg$,
N is to the left, south is to the right}

\figcaption{Isophotes of the emission line [NII]$\lambda$6583 surface
brightness distribution on the gray-scaled continuum $\lambda$6500}

\figcaption{Isolines of the surface distributions of:
{\it a} -- continuum brightness at $\lambda$5100,
{\it b} -- the absorption-line index Mgb,
{\it c} -- the absorption-line index $<\mbox{Fe}>$,
{\it d} -- the absorption-line index H$\beta$.
The continuum is calibrated in arbitrary units; the indices are
in the Lick system multiplied by 100. The cross marks the position
of continuum peak brightness}

\figcaption{Comparison of our observational data for NGC~7331
with the models of Worthey (1994) for [Mg/Fe]=0 on the diagram
(Fe5270, Mgb). The points connected by a dashed line are azimuthally
averaged and taken along the radius with the step of $1\farcs 3$.
The ages of the Worthey's (1994) models are given in billion years}

\figcaption{The age-diagnostics diagrams for NGC~7331: {\it a} --
H$\beta$ vs $< \mbox{Fe} >$, the models of Tantalo et al. (1998)
for [Mg/Fe]=+0.3
valid for the nucleus and for the bulge at $R > 3\arcsec$,
{\it b} --
H$\beta$ vs $< \mbox{Fe} >$, the models of Tantalo et al. (1998)
for [Mg/Fe]=0.0
valid for the circumnuclear disk,
{\it c} --
H$\beta$ vs [MgFe], the models of Worthey (1994) for [Mg/Fe]=0,
valid for the circumnuclear disk.
The points connected by a dashed line are azimuthally averaged and
taken along the radius with the step of $1\farcs 3$; the points
corresponding to the Fe-rich circumnuclear disk are for individual
spatial elements.
The ages of the models are given in billion years;
the metallicities for the Worthey's models are +0.50, +0.25, 0.00,
--0.22, --0.50, --1.00,--1.50, --2.00, if one takes the signs from
the right to the left, and for the models of Tantalo et al. they are
+0.4, 0.0, and -0.7}

\figcaption{Two-dimensional line-of-sight velocity fields for the
stars ({\it a}) and for the ionized gas ({\it b}). The cross marks
the position of the continuum peak brightness}

\figcaption{The direct view of the stellar LOSVD along the major axis
of NGC~7331 obtained from the long-slit data. The horizontal axis is
a velocity direction, with the full range of 2250 km/s; the vertical
axis is a spatial direction (along the slit), with the full range of
$111\arcsec$, the nucleus is at the middle}

\figcaption{Line-of-sight stellar velocity profiles along the major
axis ({\it a}) and along the minor axis ({\it b}) obtained for
NGC~7331 from the long-slit data}


\begin{thebibliography}{}

\bibitem[1989]{asz89} Afanasiev, V.L., Sil'chenko, O.K., Zasov, A.V.
   1989, \aap, 213, L9

\bibitem[Afanasiev et al. 1990]{afetal90} Afanasiev, V.L., Vlasyuk,
   V.V., Dodonov, S.N., Sil'chenko, O.K. 1990, Preprint SAO, N54
   (Nizhnij Arkhyz: Special Astrophys. Obs.)

\bibitem[Afanasiev \& Sil'chenko 1999]{we99} Afanasiev, V.L., Sil'chenko
   O.K. 1999, \aj, 117, in press

\bibitem[1972]{ak72} Arp, H.C., Kormendy, J. 1972, \apj, 178, L101

\bibitem[1998]{bag} Baggett, W.E., Baggett, S.M., Anderson, K.S.J.
   1998, \aj, 116, 1626

\bibitem[Begeman 1987]{begeman} Begeman, K. 1987, Ph.D. thesis,
   Univ. of Groningen

\bibitem[1981]{bor} Boroson, T. 1981, \apjs, 46, 177

\bibitem[1981]{bosma} Bosma, A. 1981, \aj, 86, 1791

\bibitem[1993]{betal} Bower, G.A., Richstone, D.O., Bothun, G.D.,
   Heckman, T.M. 1993, \apj, 402, 76

\bibitem[1998]{cgcg} Cardiel, N., Gorgas, J., Cenarro, J., Gonzalez,
    J.J. 1998, \aap Suppl. Ser., 127, 597

\bibitem[1999]{bhmass} Cattaneo, A., Haehnelt, M.G., Rees, M.J. 1999,
   \mnras, submitted (astro-ph/9902223)

\bibitem[1994]{crb} Cowan, J.J., Romanishin, W., Branch, D. 1994,
   \apj, 436, L139

\bibitem[1997]{dav} Davidge, T.J. 1997, \aj, 113, 985

\bibitem[Dressler \& Richstone 1988]{dr88} Dressler, A., Richstone, D.O.
   1988, \apj, 324, 701

\bibitem[1996]{ems96} Emsellem, E., Bacon, R., Monnet, G., Poulain, P.
   1996, \aap, 312, 777

\bibitem[1998]{hr} Heraudeau, P. and Simien, F. 1998, \aaps, 133, 317

\bibitem[Hughes et al. 1998]{} Hughes, S.M.G., Han, M., Hoessel, J.,
   Freedman, W.L., et al. 1998, \apj, 501, 32

\bibitem[1983]{keel} Keel, W.C. 1983, \apj, 268, 632

\bibitem[1987]{kent} Kent, S.M. 1987, \aj, 93, 816

\bibitem[Kormendy 1988]{k88} Kormendy, J. 1988, \apj, 325, 128

\bibitem[Marcelin et al. 1994]{fr} Marcelin, M., Petrosian, A.R.,
   Amram, P., Boulesteix, J. 1994, \aap, 282, 363

\bibitem[1997]{esptig} Mediavilla, E., Arribas, S., Garcia-Lorenzo, B.,
   del Burgo, C. 1997, \apj, 488, 682

\bibitem[Monnet et al. 1992]{mbe92} Monnet, G., Bacon, R., Emsellem, E.
   1992, \aap, 253, 366

\bibitem[1989]{pogge} Pogge, R.W. 1989, \apjs, 71, 433

\bibitem[1996]{espls} Prada, F., Gutierrez, C.M., Peletier, R.F.,
   McKeith, C.D. 1996, \apj, 463, L9

\bibitem[1992]{pretal} Prieto, M., Longley, D.P.T., Perez, E., Beckman,
   J.E., Varela, A.M., Cepa, J. 1992, \aaps, 93, 557

\bibitem[Rubin et al. 1965]{retal65} Rubin, V.C., Burbidge, E.M.,
   Burbidge, G.R., Crampin, D.J., Prendergast, K.H. 1965, \apj, 141,
   759

\bibitem[1961]{ha} Sandage, A. 1961, The Hubble Atlas of Galaxies
   (Washington: Carnegie Institution of Washington)

\bibitem[Sil'chenko et al. 1992]{sil92} Sil'chenko, O.K., Afanasiev,
   V.L., Vlasyuk, V.V. 1992, \azh, 69, 1121

\bibitem[1998]{smith} Smith, B.J. 1998, \apj, 500, 181

\bibitem[1998]{src} Stockdale, Ch.J., Romanishin, W., Cowan, J.J. 1998,
   \apj, 508, L33

\bibitem[1998]{tantalo} Tantalo, R., Chiosi, C., Bressan, A. 1998,
   \aap, 333, 419

\bibitem[1997]{jap} Tosaki, T., Shioya, Y. 1997, \apj, 484, 664

\bibitem[Vlasyuk 1993]{vlas} Vlasyuk, V. V. 1993, Astrofiz. issled.
    (Izv. SAO RAS) 36, 107

\bibitem[1996]{germ} von Linden, S., Reuter, H.-P., Heidt, J.,
   Wielebinski, R., Pohl, M. 1996, \aap, 315, 52

\bibitem[1994]{worth94} Worthey, G. 1994, \apjs, 95, 107

\bibitem[Worthey et al. 1994]{woretal} Worthey, G., Faber, S.M.,
   Gonzalez, J.J., Burstein, D. 1994, \apjs, 94, 687

\bibitem[Young \& Scoville 1982]{ys} Young, J., Scoville, N. 1982,
   \apj, 260, L41

\bibitem[1996]{zs96} Zasov, A.V., Sil'chenko, O.K. 1996, In: Barred
   Galaxies, Proc. of IAU Coll. 157/Eds. R. Buta, D.A. Crocker, and
   B.G. Elmegreen, ASP Conf. Ser., 91, 207


\end{thebibliography}
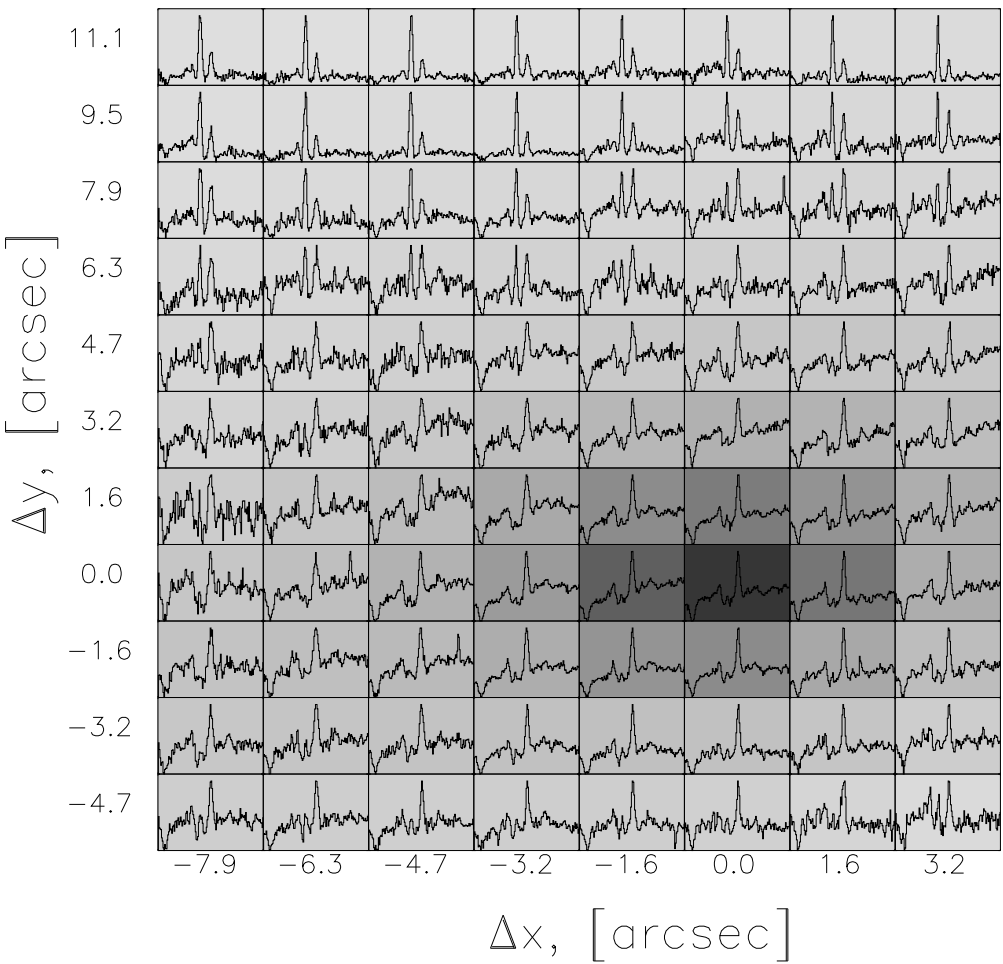
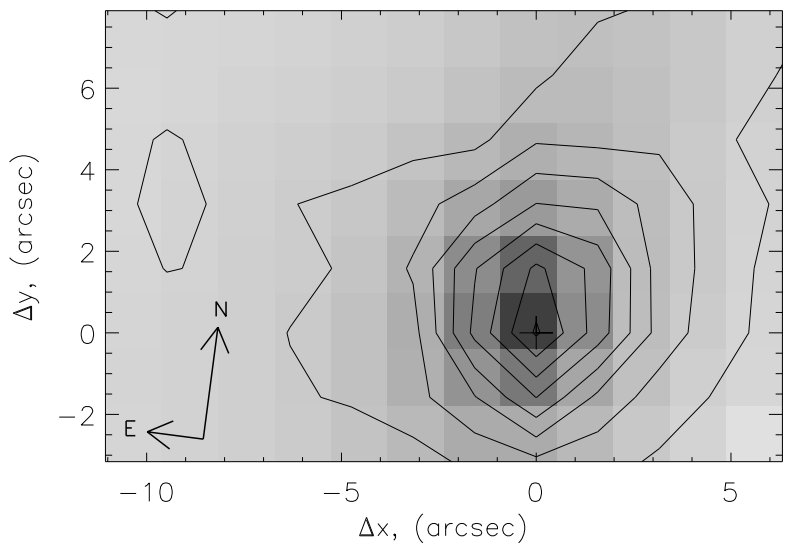
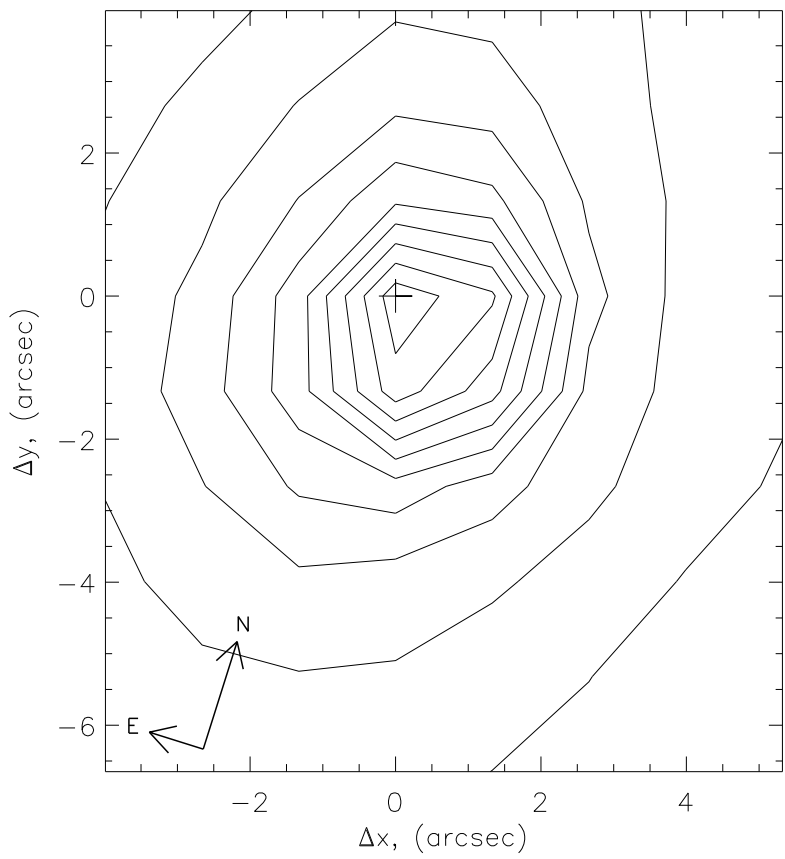
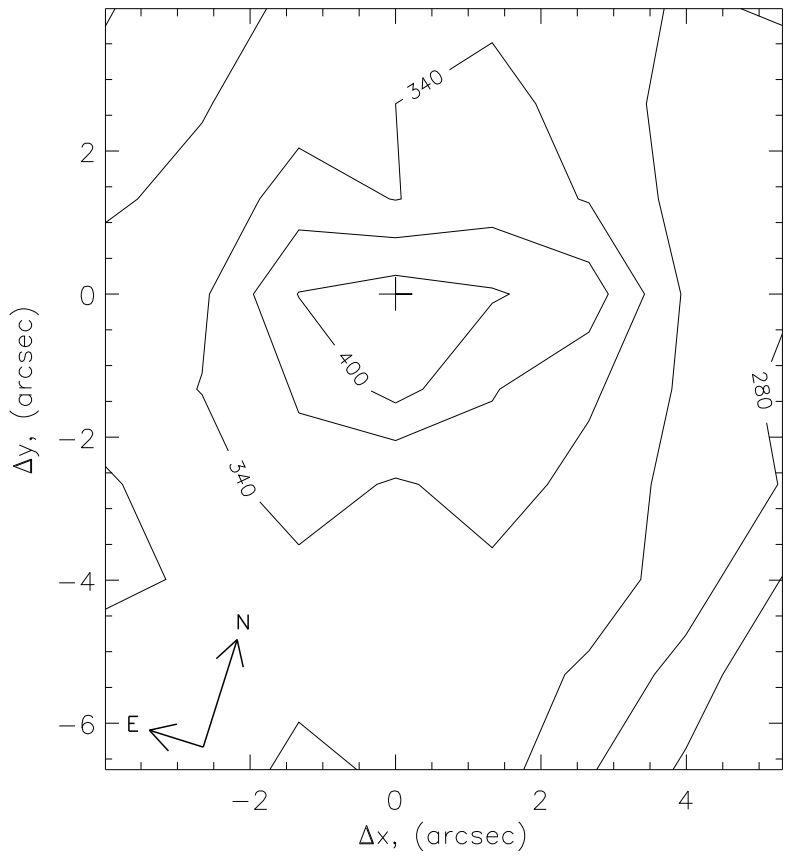
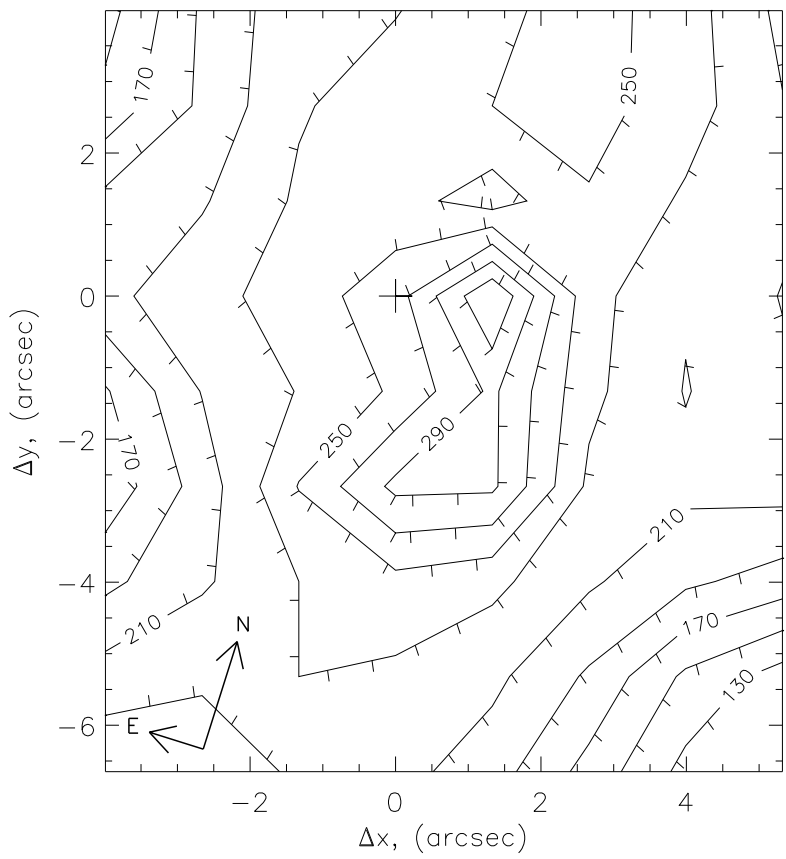
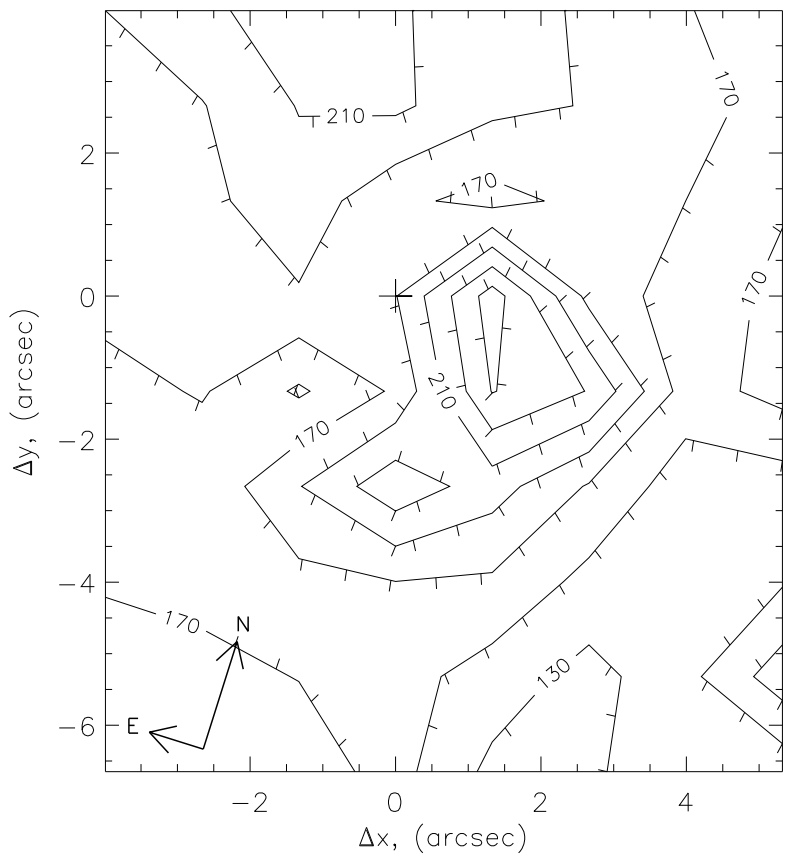
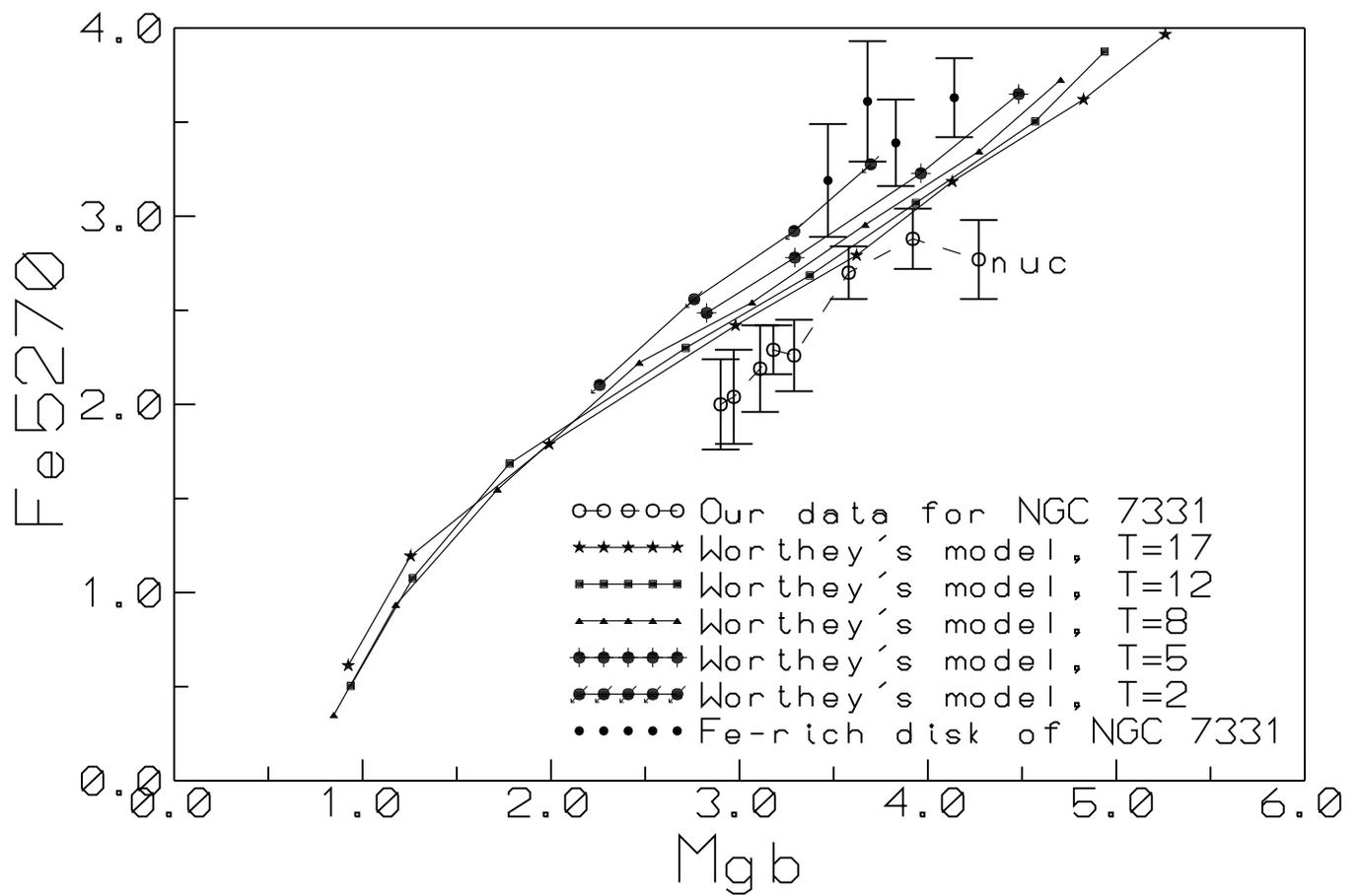
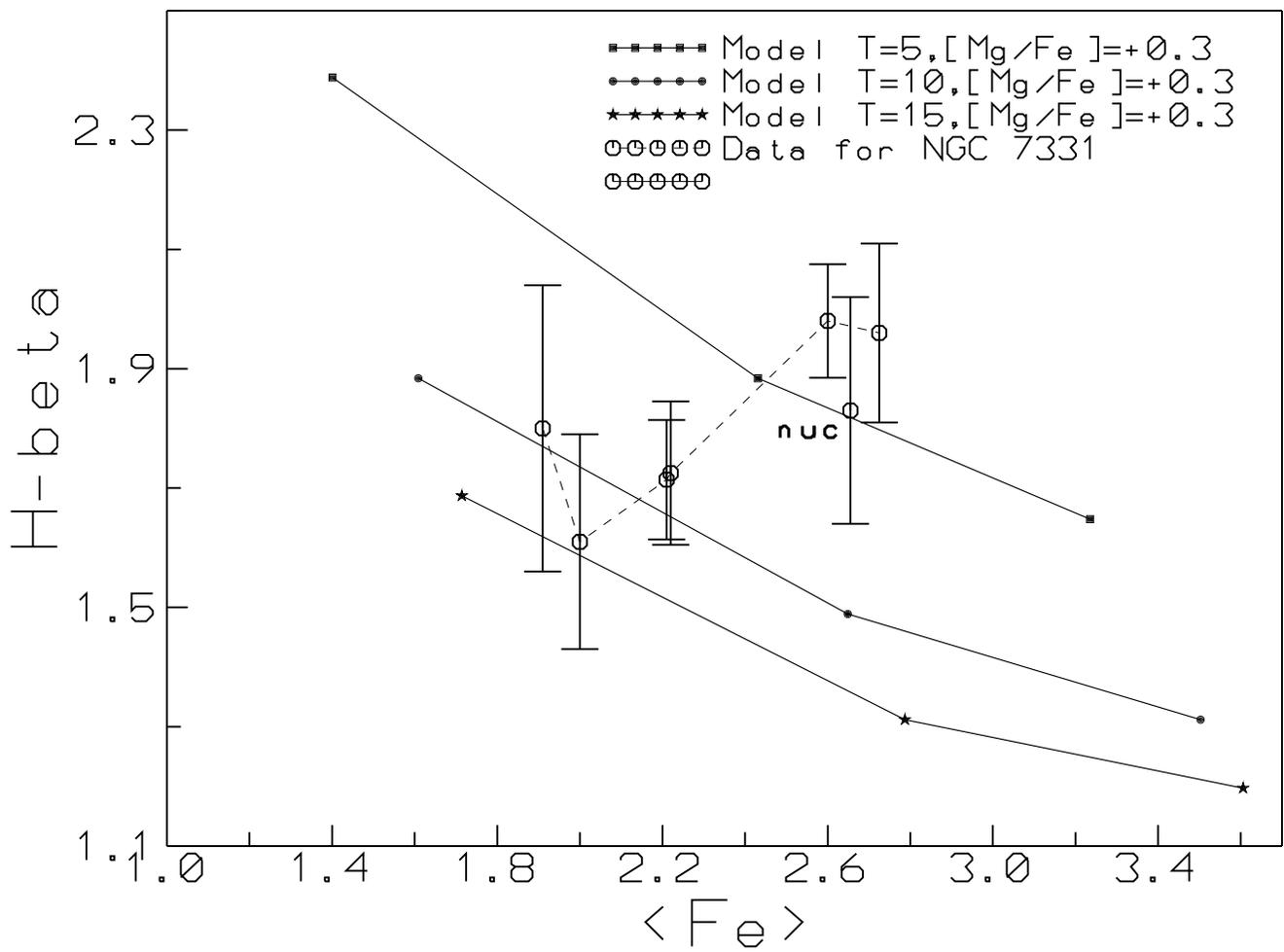
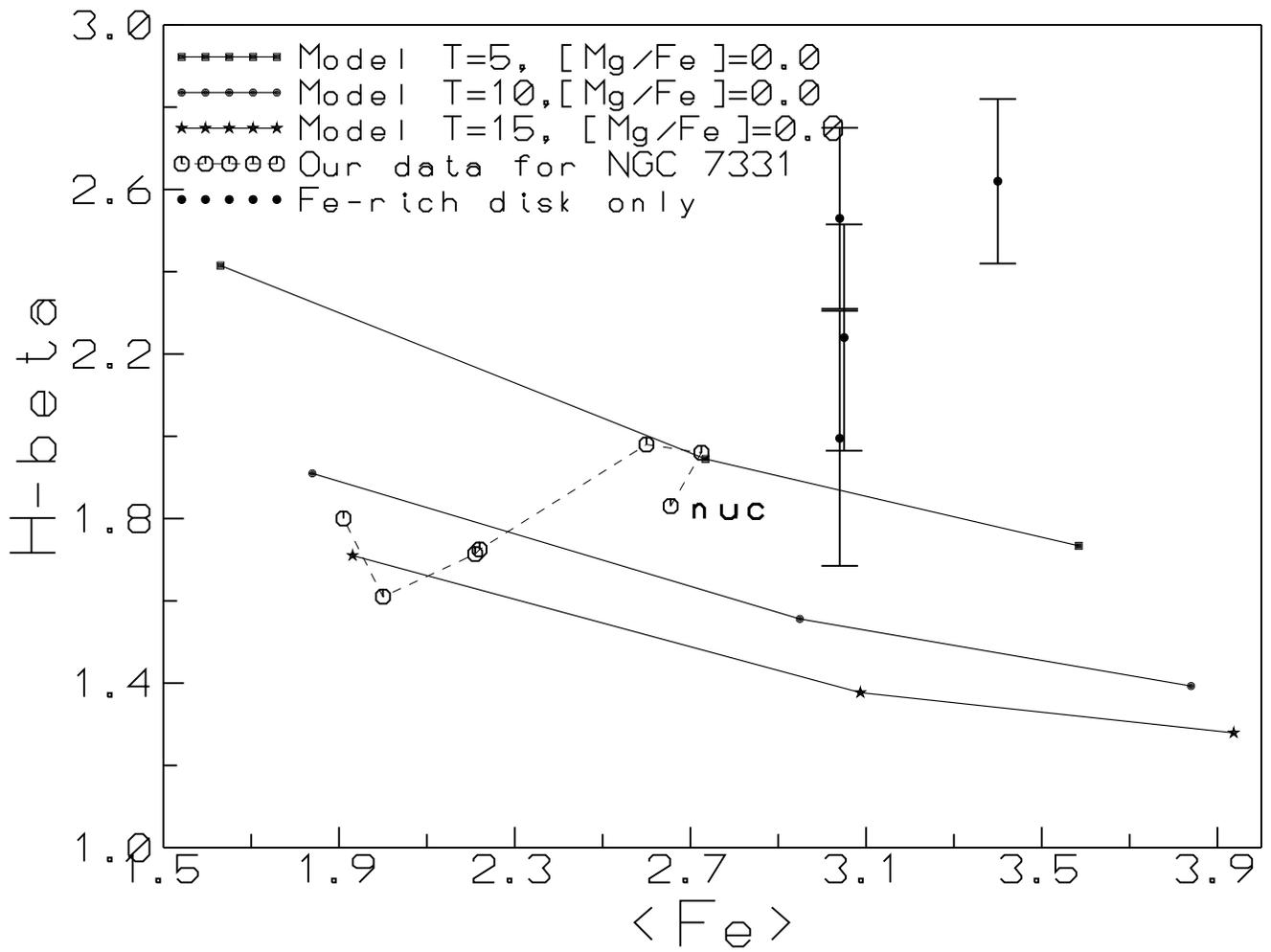
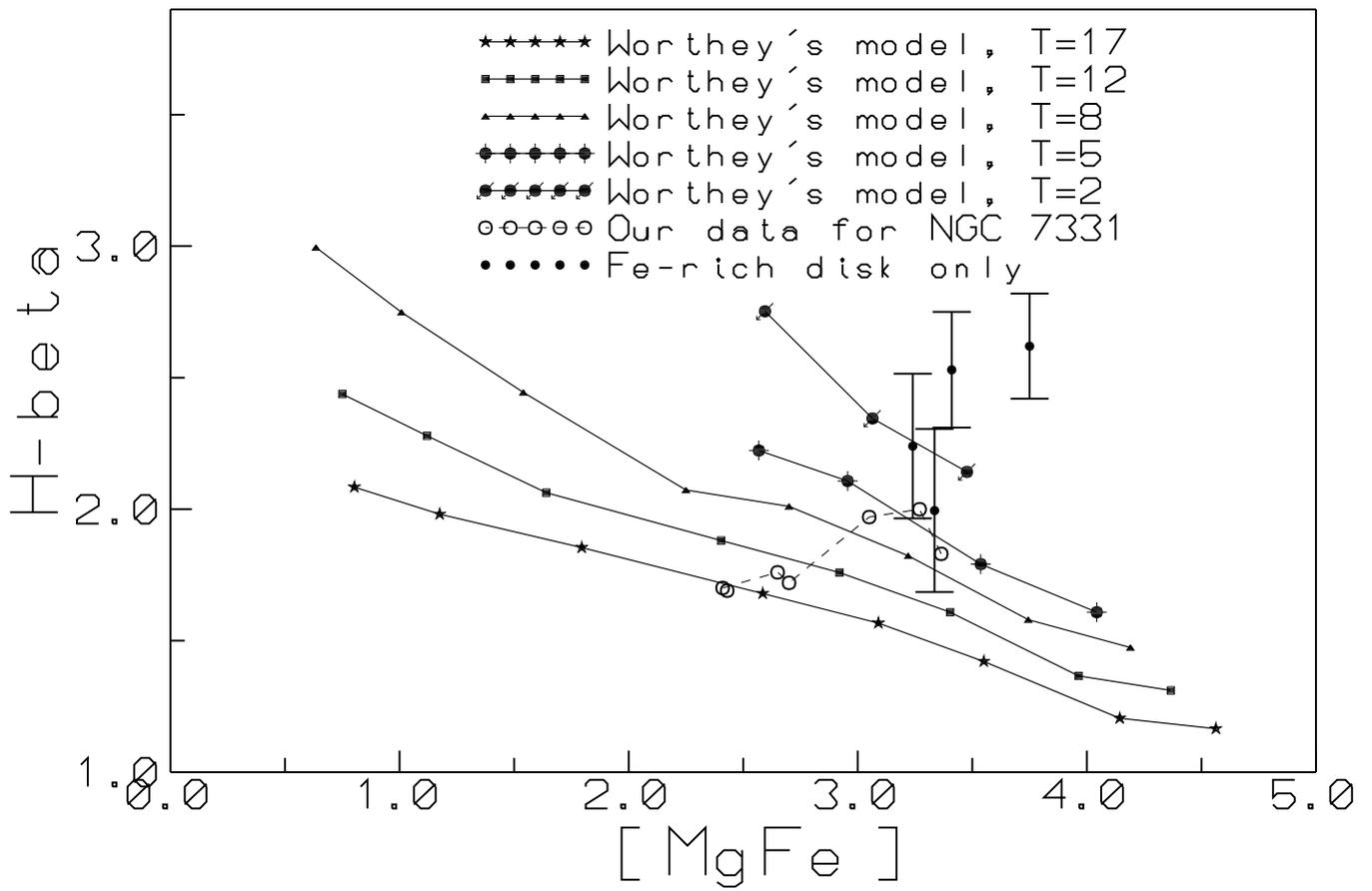
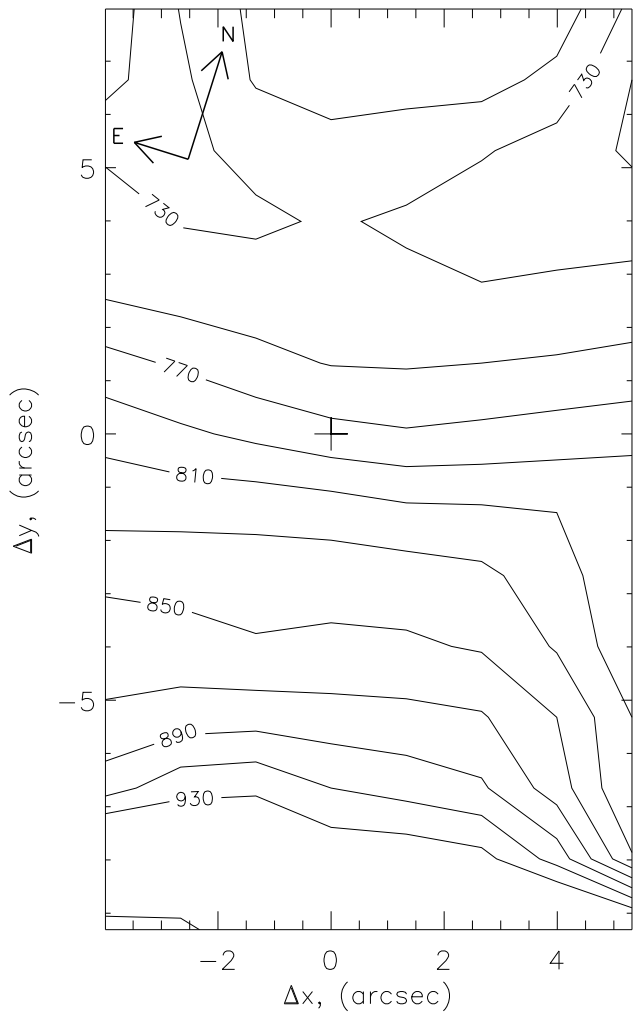
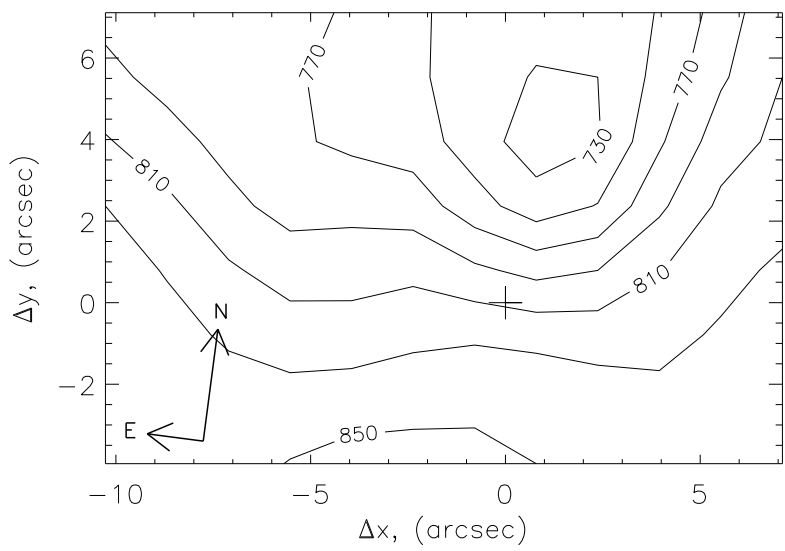
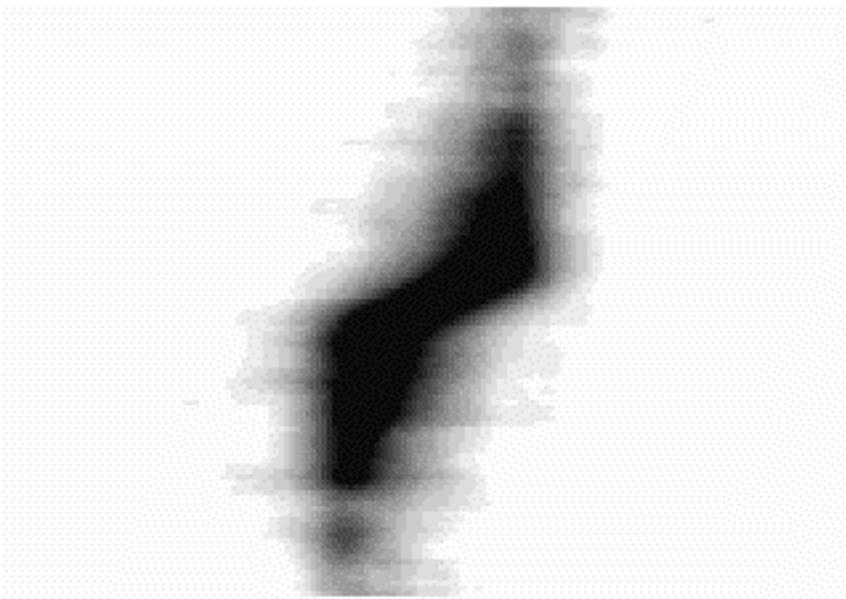
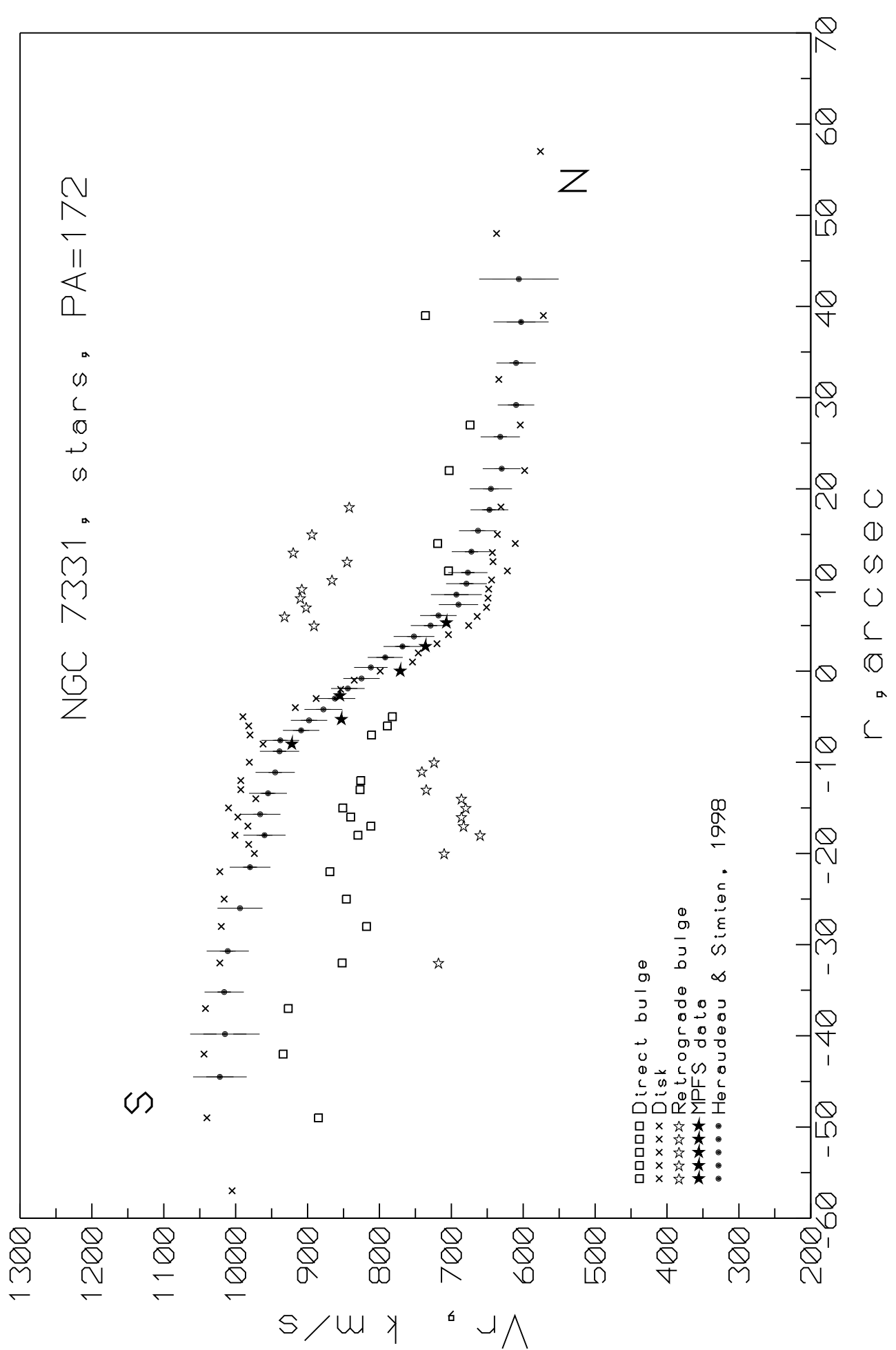
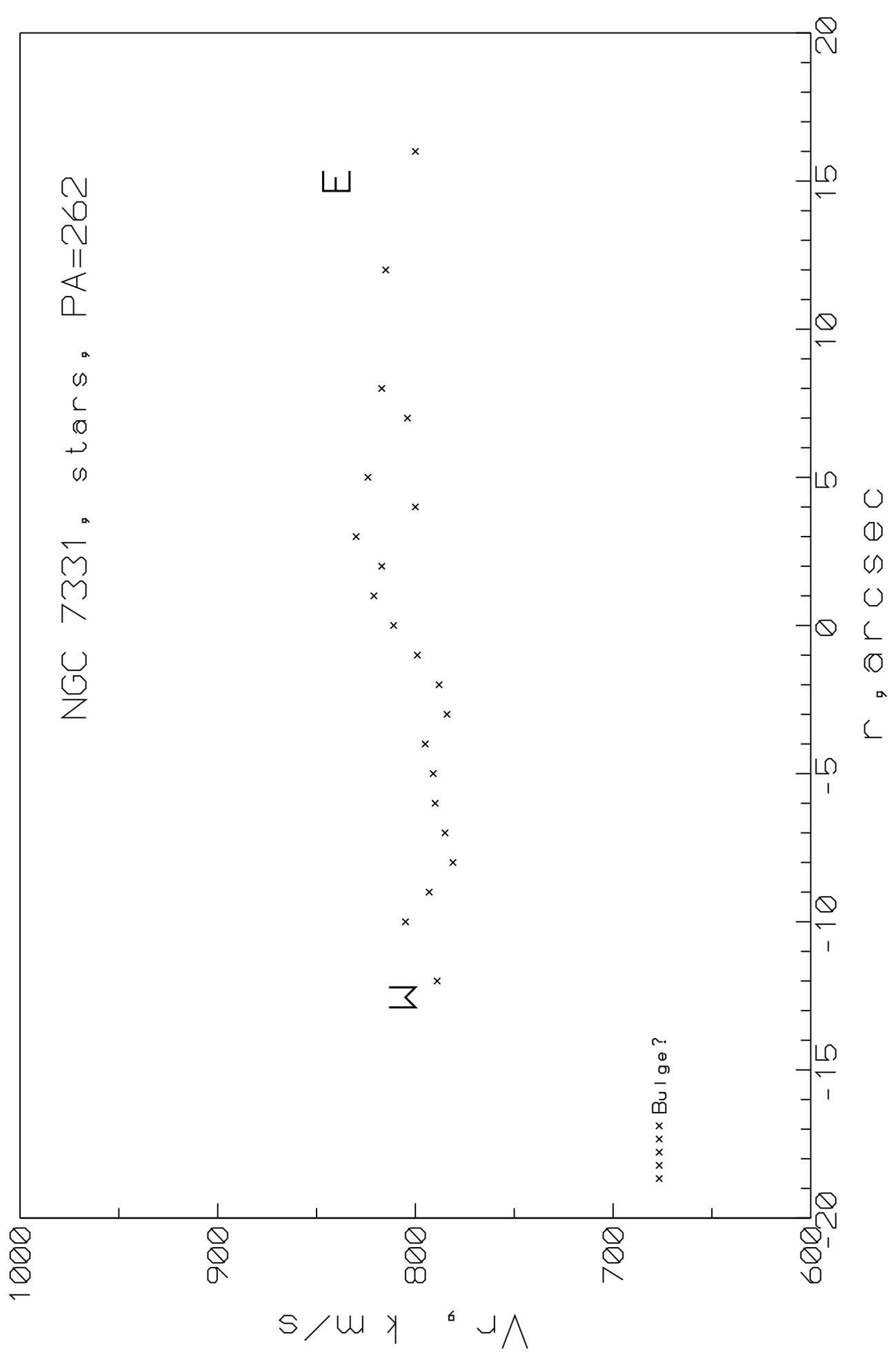
\end{document}